\documentstyle[11pt,newpasp,twoside,epsf]{article}
\newcommand{\HI}{H{\sc i}\ }
\newcommand{\Msun}{$M_{\odot}$}

\begin{document}
\title{The Disturbed ISM of the Local Group Dwarf Galaxy NGC~6822}
 \author{W.J.G.~de~Blok}
\affil{Australia Telescope National Facility, PO Box 76, Epping NSW 1710, Australia, email: edeblok@atnf.csiro.au}
\author{F. Walter}
\affil{California Institute of Technology, Astronomy Department 105-24, Pasadena, CA 91125, USA, email: fw@astro.caltech.edu}

\begin{abstract}

We present a first wide-field, high spatial and velocity resolution
map of the entire extended H{\sc i} distribution of the nearby Local
Group dwarf galaxy NGC 6822.  The observations were obtained with
the Parkes single-dish telescope and the Australia Telescope Compact Array
in mosaicing mode. NGC 6822 has an extended H{\sc i}-disk which is
shaped by the presence of numerous H{\sc i} holes and shells,
including a supergiant shell, and the effects of tidal interaction, in
the form of a tidal arm and an infalling or interacting H{\sc i}
complex.  These tidal features are not obvious in lower resolution
data, and only the proximity of NGC 6822 enables us to see them
clearly. This suggests that the importance of minor interactions in dwarf
galaxies may be larger than previously assumed.

\end{abstract}

\section{Introduction}

Dwarf irregular galaxies are ideal laboratories to study galaxy
evolution.  Their relatively simple structure, without dominant spiral
arms, bulges, and other complicating properties make it less difficult
to disentangle the various physical processes occuring in these
galaxies. Their low metallicities and gas-richness suggest that they
are still in an early stage of their conversion from gas into stars,
and that they could therefore provide information about galaxy
evolution as it took place in normal galaxies at earlier epochs.

NGC\,6822 is one of the closest gas-rich dwarfs in our Local Group and
is a proto-typical dwarf irregular.  Discovered by Barnard (1884), it
was studied in detail by Hubble (1925) who showed that its distance
placed it well outside our Milky Way galaxy, making NGC 6822 the first
galaxy to be recognized as an ``extra-galactic'' system.

No nearby companions are known (Mateo 1998, van den Bergh 1999).  The
galaxy is at a distance of $490\pm 40$ kpc (Mateo 1998), giving a
linear resolution of $1'' = 2.5$ pc. Its absolute magnitude is
$M_B=-15.85$ (Hodge et al.\ 1991) making it $\sim 1.5$ mag fainter
than the SMC.  The optical diameter $d_{25}$ is $18.3'$ (2.7 kpc). Its
optical appearance is dominated by a Magellanic bar and star forming
regions.  The metallicity is low: the oxygen abundance is $12 + \log
O/H = 8.3 \pm 0.1$ (Skillman et al.\ 1989) and its stellar metallicity
is 0.004 $Z_{\odot}$ (Gallart et al.\ 1996a). 

In the past, NGC\,6822 has already been the subject of many studies
investigating its stellar population, sites of star formation and star
formation history.  (e.g.\ Gallart et al.\ 1996abc, Hodge et al.\
1991).  Due to its proximity such studies are already feasible using
ground-based optical telescopes. Other studies have investigated
possible relations between star formation and molecular content (e.g.\
Petitpas \& Wilson 1998, Israel 1997) and the galaxy has been
extensively studied in X-rays using the {\sc rosat} and {\sc einstein}
satelites (e.g.\ Eskridge \& White 1997).

While its low galactic latitude ($b = -18.4^{\circ}$) is a nuisance
for optical studies (because of the high stellar density, foreground
extinction, reddening [$E(B-V) = 0.24 \pm 0.03$; Gallart et al.\
1996a]), its low recession velocity of $-55$ km\,s$^{-1}$ has proven
to be a potential problem for 21-cm neutral hydrogen \HI observations
as the velocities of the local Galactic emission partly overlap those
of NGC\,6822. This confusion, as well as NGC\,6822's large size, may
be the reason why to date only surprisingly few \HI studies of this
galaxy have appeared in the literature, even though its inclination,
gas-richness, isolation and proximity make it an ideal candidate for
studying the rotation curve, dark matter content and relation between
the stellar population and gaseous ISM.

The earliest \HI study of NGC\,6822 is that by Volders \& H\"ogbom
(1961) who used the 25m Dwingeloo telescope to probe NGC\,6822 and its
environs.  They found evidence for a large amount of neutral hydrogen
($1.5 \times 10^8$ \Msun) rotating in ordered fashion around the
optical center.  Later studies by Davies (1970) and Roberts (1970),
confirmed these results.  The first interferometer study was done by
Gottesman \& Weliachew (1977; G\&W hereafter) who observed the optical
centre in HI using the Owens Valley Radio Observatory interferometer.
Their observations, with a resolution of $2.3'$ (330 pc) already
showed a highly structured ISM as was later confirmed by Brandenburg
\& Skillman (1998) (see also Hodge et al.\ 1991).

To trace the full extent of NGC\,6822's neutral hydrogen distribution
we present first results of new wide-field and
high-velocity-resolution observations obtained with the narrowband
filterbank and the multibeam system at the Parkes Telescope and
multiple configurations of the Compact Array synthesis radio
telescope.  The high velocity resolution allows us to separate
Galactic from galaxy emission and therefore gives us the clearest view
yet of the \HI in NGC\,6822.

\section{Single Dish data}

NGC\,6822 was observed on 16 December 1998 using the ATNF Parkes
Telescope in Australia utilising the narrow-band back-end system and
the multi-beam system.  We used a bandwidth of 4 MHz and channel width
of 0.8 km\,s$^{-1}$. The telescope beam size is 14.4$'$ at 21\,cm.  An
area of $4 \times 4$ degrees was observed.  The individual
bandpass-corrected single-beam, single-polarization spectra were
gridded into a data cube.  Because of the small amount of smoothing
that is intrinsic to the gridding process the effective beam size
ended up being 16.7$'$ (2.5\,kpc).  As expected, confusion with the
Galaxy was present in a number of channels --- in these we were able
to isolate the NGC\,6822 signal by determining in each channel map the
median flux in a ring surrounding the NGC\,6822 signal and subtracting
that value from the channel map.  

\begin{figure}
\plotone{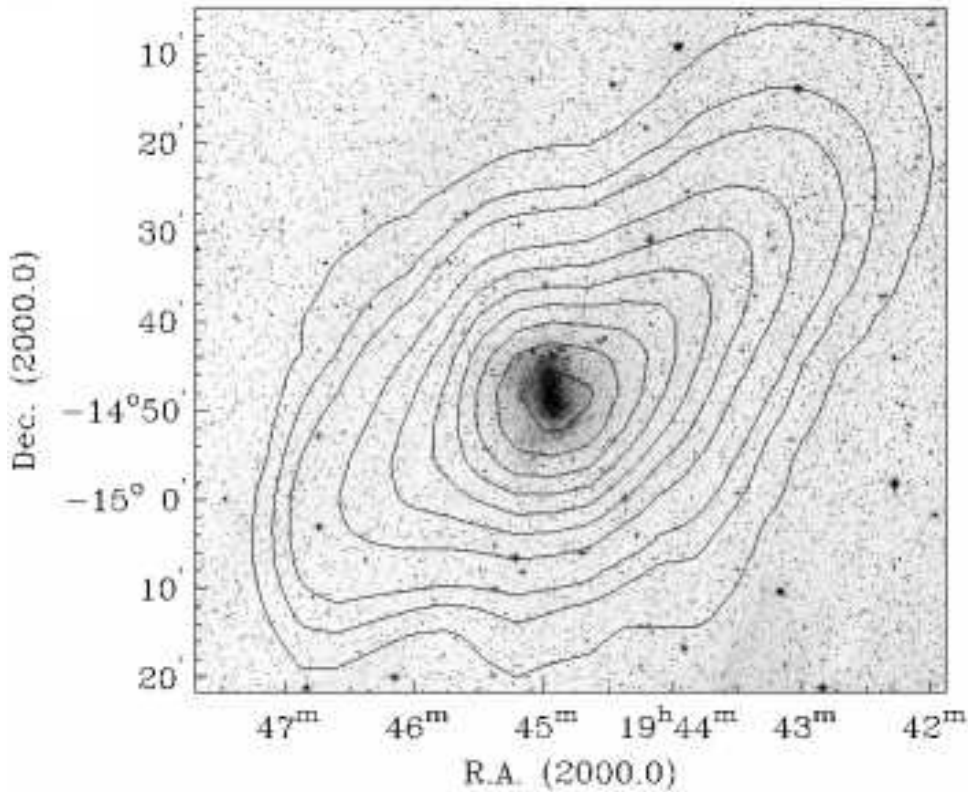}
\caption{Total \HI column density map of NGC\,6822 (contours) overlaid on 
an optical image from the DSS (grayscales). Contours are 0.2, 0.5, 1,
2, 3, 4, 5, 6, 7, 8 $\times 10^{20}$ atoms cm$^{-2}$. The effective
beam size is $16.7'$. The \HI distribution is clearly more extended
than the optical distribution.
\label{nb_himap}}
\end{figure}

A total \HI column density map is shown in Fig.~\ref{nb_himap}
overlaid on an image obtained from the Digital Sky Survey (DSS). Note
that the \HI is much more extended than the optical galaxy.  Even when
taking the large beam size of $16'$ (2.5\,kpc) into account the extent
of \HI along the major axis is still more than a degree (9\,kpc).
Compared to the optical $d_{25}$ diameter the H{\sc i} disk is about
three times more extended. The disk looks asymmetrical with the SE
(receding) side less extended than the NW (approaching) side. This
suggests that the \HI disk is intrinsically asymmetric, as hinted at
by the G\&W(1977) observations of the inner $20'$.
We derive a total \HI-mass of $(1.3 \pm 0.1) \times 10^8$ \Msun\ from
these data.

A full dynamical analysis of NGC\,6822 is beyond the scope of this
paper. This would require a detailed analysis of the rotation curve at
higher resolution as well as a mass decomposition using the various
mass components.  We can however still estimate the dark matter
content of NGC\,6822. This only depends on $V_{\rm max}$, which is a
well-determined quantity.  We find $V_{\rm max} = 57$ km\,s$^{-1}$ at
$R_{\rm max}=5.7$ kpc. This yields a dynamical mass within that radius
of $M_{\rm dyn} = 4.3 \times 10^9$~\Msun .

For computing the visible mass we use the \HI mass derived above,
multiplied by 1.4 to take the helium contribution into account.  For
the stellar mass we use the absolute $B$-band magnitude\footnote{
Unfortunately $M_B$ is not very well constrained. Hodge et al (1991)
use $M_B = - 15.85$; Gallart et al.\ (1996a) quote $M_B = -14.1$; G\&W
(1977) use $M_B = -15.1$. In the following we have used the mean of
these three numbers.}  and assume a stellar mass-to-light ratio
$M/L^B_\star = 1$, which is not an unreasonable value for dwarf
galaxies.  This yields $M_{HI} / L_B = 0.7$, $M_{\rm dyn}/ L_B = 23$,
$M_{\rm dyn}/ M_{\rm vis} = 11$.  It is clear that dark matter
dominates the dynamics of NGC\,6822.

The global properties of NGC 6822 are those of a typical dwarf
irregular galaxy. The single dish data presented here have the same
{\it linear} resolution as those made of a dwarf galaxy at 30 Mpc
using an instrument like the VLA, WSRT or ATCA. We are now in the
position to use follow-up high-resolution ATCA data to ``zoom in'' on
NGC 6822 and see what the low resolution is hiding from us.

\section{Compact Array Observations}

NGC\,6822 was observed with the Australia Telescope Compact Array for
$15 \times 12$ hours in its 375, 750D, 1.5A, 6A and 6D configurations
over the period from June 1999 to March 2000. A total of 8 pointings
was observed covering the entire H{\sc i} extent of the galaxy.  We
used a bandwidth of 4 MHz, divided into 1024 channels, giving a
channel separation of 0.8 km s$^{-1}$.  In this paper we will restrict
ourselves to the data set that does not include the 6km
configurations. Also this data is not yet corrected for missing short
spacings. The clean synthesized beam of this medium resolution data
set measures $89'' \times 24''$ ($222 \times 60$ pc).

\begin{figure}
\plotfiddle{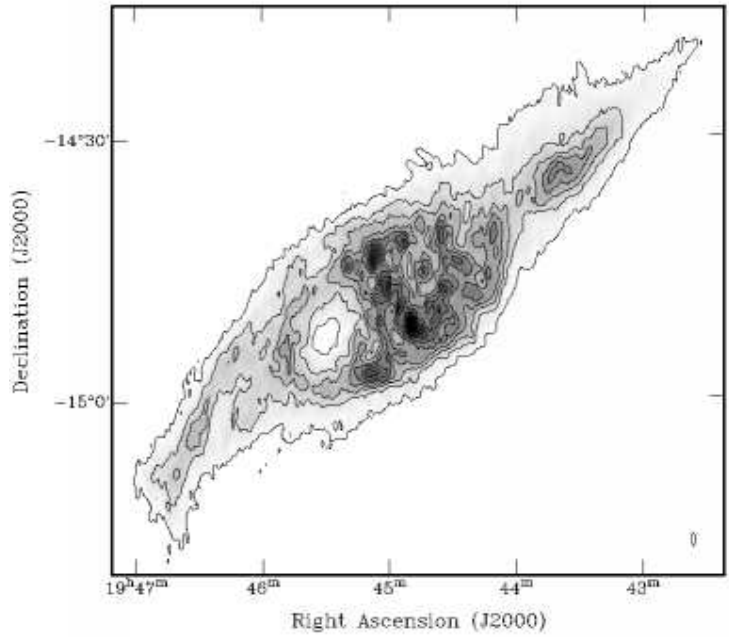}{0.3\vsize}{-90}{45}{45}{-150}{300}
\plotfiddle{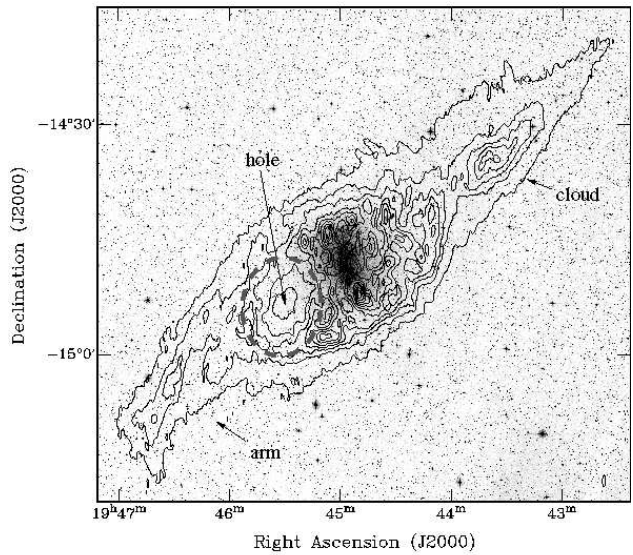}{0.4\vsize}{0}{120}{120}{-380}{-350}
\caption{Top: Integrated H{\sc i} 
column density map. Contours are $(1, 4, 7, 10, ..., 31) \times
10^{20}$ cm$^{-2}$. The beam of $89'' \times 24''$ is indicated in the
lower-right corner.  Bottom: Integrated H{\sc i} column density
contours overlaid on optical image from the DSS. Contour values are
same as in left panel. Note how the inner edge of the hole is traced
by the optical emission, notably in the south. The three features of
hole, cloud and arm as discussed in the text are indicated. The
outline of the hole is shown by the grey dashed ellipse.}
\end{figure}

Figure 2 shows the integrated H{\sc i} column density map as derived
from our data. Also shown is the column density map overplotted on an
optical DSS image.  The H{\sc i} is much more extended than the
optical distribution, extending out to galactocentric radii of 40
arcmin ($\sim 6$ kpc). We find a total \HI mass of $1.1 \times 10^8$
\Msun.  A comparison with the single dish Parkes \HI mass of $(1.3 \pm
0.1) \times 10^8$ \Msun\ shows that we are missing only a small
fraction of the flux due to missing short spacings.

{\section{Features in the neutral ISM}

A lot of structure in the form of holes and shells is present in the
H{\sc i} disk. The three striking features are apparent that we will
discuss briefly: these are the hole and arm in the SE, and the cloud
in the NW. For a more extensive discussion of these features we refer to
de Blok \& Walter (2000).

{\bf A supergiant HI hole}\  In the SE a giant H{\sc i} hole or shell is
dominating the appearance of the galaxy.  Its angular size size is
$14'\times 10'$ (as indicated in Fig.~1), corresponding to $2.0\times
1.4$ kpc, measured at a column density level of $10^{21}$
cm$^{-2}$ (see also Brandenburg \& Skillman 1998).

In the standard picture, H{\sc i} shells and supershells are caused by
the stellar winds of the most massive stars in a cluster as well as
subsequent SN explosions (for reviews see Kulkarni
\& Heiles 1988, van der Hulst 1996 and Brinks \& Walter 1998).
If the hole was indeed created by past massive star formation, we can
derive some approximate numbers regarding the energies and ages
involved. If we assume that the expansion velocity of the hole has
reached values similar to the dispersion of the ambient ISM in
NGC\,6822 ($\sim 7$km s$^{-1}$), we derive a kinematic age of 130
Myr. The kinematic age is an upper limit for the actual age since the
shell was presumably expanding more rapidly in the past. Using
Chevalier's equation (Chevalier 1974), we derive an energy of
$10^{53}$\ erg needed to create the H{\sc i} shell, equivalent to 100
Type II supernovae (using $V_{exp} = \sigma_{gas} =7$\ km s$^{-1}$,
$n_{HI} = 0.1$\ cm$^{-3}$ and $r_{hole}=850$\ pc). Note that it is not
necessary that these supernovae go off at the same time, which would
need a massive parent cluster. It is more likely that many sequential
events created a big hole by superposition.

{\bf An H{\sc i} companion?}
The extreme NW of the galaxy contains an H{\sc i} complex with a
morphology different from the rest of the galaxy.  Unfortunately, it
is difficult to tell whether it actually belongs to the main disk of
NGC\,6822 or whether it is a companion which happens to be at a
similar heliocentric velocity. The H{\sc i} mass of the NW complex is
$\sim 1.4 \times 10^7\ M_{\odot}$, i.e.\ $\sim 10\%$ of the total
H{\sc i} mass of the NGC~6822-system. 

That the NW complex is a separate system is supported by the asymmetry
in the H{\sc i} disk of NGC 6822.  The NW half contains 20\% more
H{\sc i} than the SE half (a difference of $\sim 1.2\times 10^7\
M_{\odot}$) as measured with respect to a minor axis passing through
the geometrical center.  Assuming that the disk of NGC 6822 is
symmetric to begin with, this asymmetry can be explained by assuming
that the NW cloud (with a mass of $\sim 1.4\times 10^7\ M_{\odot}$) is
a separate system contributing to the mass in the NW half.

{\bf Tidal effects}\ A third interesting feature, resembling a tidal
arm, is visible in the SE. We think it is unlikely that it is a
conventional spiral arm, due to the absence of star formation in this
part of the galaxy, the absence of any spiral structure in the optical
and the inner H{\sc i} disk, and the asymmetric H{\sc i} morphology in
the NW and SE.  Whether the material in this arm was stripped off
NGC\,6822's main disk or belonged to an interaction partner is
difficult to tell based on our data. Future numerical simulations may
shed light on this situation.  As NGC 6822 is isolated, at the
outskirts of the Local Group, and not associated with any of the
subgroups, it is unlikely that it interacted with any of the known
Local Group galaxies.  One important caveat is, however, that a lot of
Galactic emission and HVC structures are present in this region of the
sky between heliocentric velocities $+25$ and $-15$ km s$^{-1}$. This
velocity range coincides with that of the SE arm. It is very well
possible that possible companions of NGC\,6822 are hidden in the
strong galactic emission.

The most likely possibility though is that the NW cloud is the
interaction partner. An upper limit to the time scale for this
encounter is of the order of half the rotation period at the radial
distance of the cloud (i.e., the time to move from SE to NW), which is
$3 \times 10^8$ yr.  A rough estimate for the timescale can also be
derived from the tidal feature itself: the arm measures some $20'$
which corresponds to about 2.8 kpc. It is difficult to estimate what
the velocity of the arm is with respect to the galaxy.  The data
suggests a value between 10 and 30 km s$^{-1}$. Using 20 km s$^{-1}$
we then derive a kinematic age of $1.4
\times 10^8$ yr, but any number between 100 and 200 Myr is probably
reasonable.  The interaction described here was a minor one, as it did
not result in a large starburst and the ejection of large amounts of
gas. Dwarf galaxies can apparently undergo such minor interactions
without a noticeable effect on their {\it global} properties. It is
the small distance to NGC 6822 that enables us to study this process
in so much detail. The Parkes single-dish maps presented in Sect.~2
and low resolution maps made using data from only the 375m ATCA
configuration show that if NGC 6822 had been a factor of $\ga 5$
further away, it would have been impossible to distinguish the NW
complex from the main body, nor would the tidal arm and the hole have
been obvious.  The number of minor interactions in dwarf galaxies may
therefore be much larger than one would guess on the basis of low to
medium resolution H{\sc i} observations of more distant galaxies.

\section{Summary}

NGC 6822 is one of the very few dwarf systems in the local universe
that allows such a detailed study of its ISM and stellar population.
We have presented low and high resolution data showing that NGC 6822
is not a quiescent non-interacting dwarf galaxy, but that it is
undergoing an interaction with a possible companion.  The timescale
for this interaction is of order $10^8$ years, comparable to the
timescale for formation of the giant hole, suggesting that these two
events are related. The passage of the companion may have been the
trigger for star formation so far out in the disk.  Similar timescales
have been derived from optical studies: Hodge (1980) finds evidence
for an enhancement in star formation between 75 and 100 Myr ago, while
the extensive study by Gallart et al.\ (1996c) shows that the SFR in
NGC 6822 increased by a factor 2 to 6 (depending spatial position)
between 100 and 200 Myr ago.  Follow-up H{\sc i}, optical and infrared
observations (allowing stellar population studies) currently in
progress and the added benefit of very high resolution in the complete
H{\sc i} dataset (beam of $6'' = 15$ pc), will enable us to study the
dynamics of the ISM in NGC 6822 from pc to kpc scales.

\acknowledgements We wish to thank Lister Staveley-Smith for his help in
obtaining the Parkes data. FW acknowledges NSF grant AST 96-13717.
The Australia Telescope is
funded by the Commonwealth of Australia for operation as a National
Facility managed by CSIRO.  This research has made use of the
NASA/IPAC Extragalactic Database (NED) which is operated by JPL,
Caltech, under contract with NASA, NASA's Astrophysical Data System
Abstract Service (ADS), and NASA's SkyView.

\end{document}